\documentclass[a4paper,11pt]{article}
\pdfoutput=1 

\usepackage{jinstpub} 
\usepackage{subcaption}
\usepackage{upgreek}
\usepackage{doi}
\usepackage{lineno}
\usepackage[export]{adjustbox}

\title{Characterisation of analogue front end and time walk in CMOS active pixel sensor}

\author[a,1]{B. Hiti\note{Corresponding author.}}
\author[a]{\hspace{-0.15cm}, V. Cindro}
\author[a]{\hspace{-0.15cm}, A. Gori\v{s}ek}
\author[b*]{\hspace{-0.15cm}, M. Franks}
\author[c]{\hspace{-0.15cm}, R. Marco-Hern\'{a}ndez}
\author[a]{\hspace{-0.15cm}, G. Kramberger}
\author[a]{\hspace{-0.15cm}, I. Mandi\'{c}}
\author[a,d]{\hspace{-0.15cm}, M. Miku\v{z}}
\author[b]{\hspace{-0.15cm}, S. Powell}
\author[e]{\hspace{-0.15cm}, H. Steininger}
\author[b]{\hspace{-0.15cm}, E. Vilella}
\author[a]{\hspace{-0.15cm}, M. Zavrtanik}
\author[b]{\hspace{-0.15cm}, C. Zhang}

\affiliation[a]{Jo\v{z}ef Stefan Institute, Ljubljana, Slovenia}
\affiliation[b]{University of Liverpool, Department of Physics, Liverpool, UK}
\affiliation[c]{IFIC (CSIC-UV), Valencia, Spain}
\affiliation[d]{University of Ljubljana, Faculty of Mathematics and Physics, Ljubljana, Slovenia}
\affiliation[e]{HEPHY, Vienna, Austria}
\affiliation[*]{now FBK, Trento, Italy}

\emailAdd{bojan.hiti@ijs.si}
\arxivnumber{}

\abstract{
In this work we investigated a method to determine time walk in an active silicon pixel sensor prototype using Edge-TCT with infrared laser charge injection. Samples were investigated before and after neutron irradiation to $5\e{14}\neqcm$. Threshold, noise and calibration of the analogue front end were determined with external charge injection. A spatially sensitive measurement of collected charge and time walk was carried out with Edge-TCT, showing a uniform charge collection and output delay in pixel centre. On pixel edges charge sharing was observed due to finite beam width resulting in smaller signals and larger output delay. Time walk below $25\up{ns}$ was observed for charge above $2000\el$ at a threshold above the noise level. Time walk measurement with external charge injection yielded identical results.
}


\newcommand{\up}[1]{\ensuremath{\,\mathrm{{#1}}}}
\newcommand{\neqcm}{\ensuremath{\,\mathrm{n}_\mathrm{eq}/\mathrm{cm^2}}}
\newcommand{\mum}{\ensuremath{\,\upmu\mathrm{m}}}
\newcommand{\el}{\ensuremath{\,\mathrm{e}^-}}
\newcommand{\ohmcm}{\ensuremath\,\Omega\,\mathrm{cm}}
\newcommand{\e}[1]{\ensuremath \cdot10^{#1}}

\begin{document}
\maketitle
\flushbottom


\section{Introduction}
Active silicon tracking detectors with fully integrated readout electronics manufactured in a large volume industrial CMOS process are a potential alternative for hybrid silicon detectors at future particle colliders, promising similar performance at a simplified production and a smaller amount of material in the tracking volume.
In recent years several prototypes capable of withstanding high radiation fields of the order of $10^{15}\neqcm$ have been developed \cite{atlaspix, monopix, malta}. These designs, called \textit{depleted monolithic active pixel sensors} (DMAPS), feature a depleted sensing layer providing a drift based, fast and radiation tolerant charge collection. 

The efficiency of charge collection after irradiation in these sensors is determined by two main factors. The first factor is the degradation of charge collection due to displacement damage in silicon bulk caused by non-ionising energy losses (NIEL), which leads to charge trapping and reduced depletion depth at a given bias voltage due to build up of space charge. These effects have been evaluated in numerous studies with DMAPS \cite{affolder, xfab, cavallaro, lf_igor, chess2}.
Another important requirement for efficient charge collection is the so called in-time efficiency, which describes the ability of the detector to correctly match hits with the original collision event. In the LHC environment particle collisions occur every $25\up{ns}$ and only hits resolved within this time window are considered as signals. 
The factors limiting the in-time efficiency are the speed of charge collection from the bulk and the time walk of the electronics, which describes the response delay spread for varying size signals crossing a fixed threshold level. 

Typically the time walk of a chip is determined from the output delay after injecting a varying test charge directly into the front end electronics. This method does not provide in-pixel positional sensitivity which is usually determined by a complementary measurement of particle detection efficiency in a test beam (such as \cite{monopix_tb}). 
In this work we investigated a new approach to evaluate spatial dependence of in-time efficiency using Edge-TCT method, which employs a focused laser beam to inject charge into different parts of the pixel. This method can be carried out in a laboratory environment and can potentially simplify sample characterisation. 
This study was introduced in the scope of characterising the properties of the analogue front end in an active silicon pixel detector prototype which is also described in this paper.

\section{Samples and measurement setup}

The active silicon pixel detector prototype investigated in this work has been developed within the CERN RD50 collaboration which aims to evaluate different aspects of radiation tolerance of industrial CMOS processes \cite{ricardo}.
The investigated sample called RD50-MPW2 has been manufactured in a $150\up{nm}$ CMOS process by LFoundry on a $280\mum$ thick p-type substrate with an initial resistivity of $1900\ohmcm$. The bias voltage for sensor depletion is applied to dedicated p-type rings surrounding each pixel on the top side of the chip and reaches the breakdown value at $120\up{V}$. 
The back side of the substrate is not processed. The electric field configuration resulting from the described biasing scheme may lead to a reduced charge collection efficiency after irradiation \cite{lf_igor}. However, the investigated response delay at a given collected charge is not affected. Detectors for practical applications will have metallised back plane to avoid this feature.

\begin{figure}%
\centering
\begin{subfigure}[c]{0.495\textwidth}
\includegraphics[width=\columnwidth]{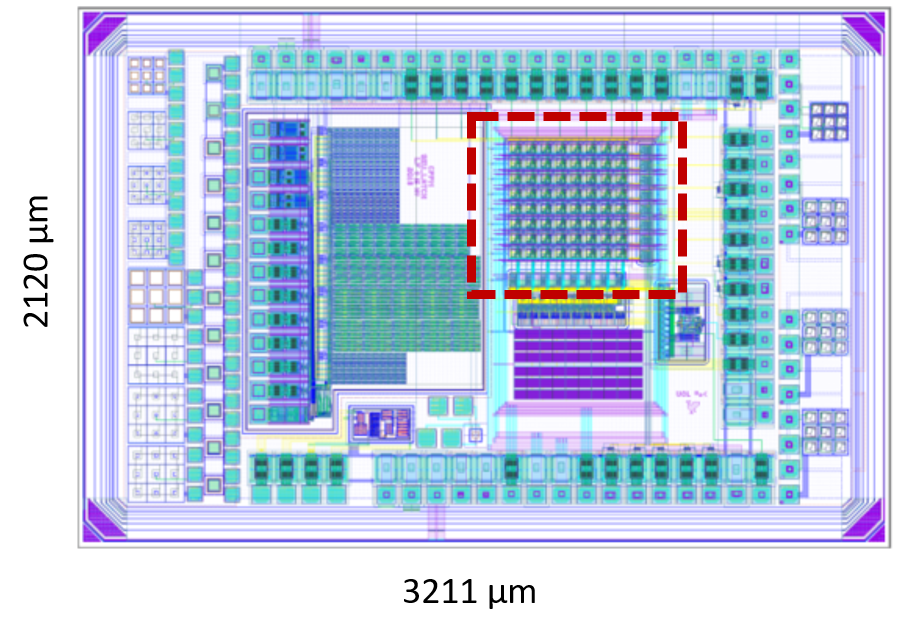}%
\caption{}
\label{fig:chip}
\end{subfigure}
\hfill
\begin{subfigure}[c]{0.495\textwidth}
\includegraphics[width=\columnwidth]{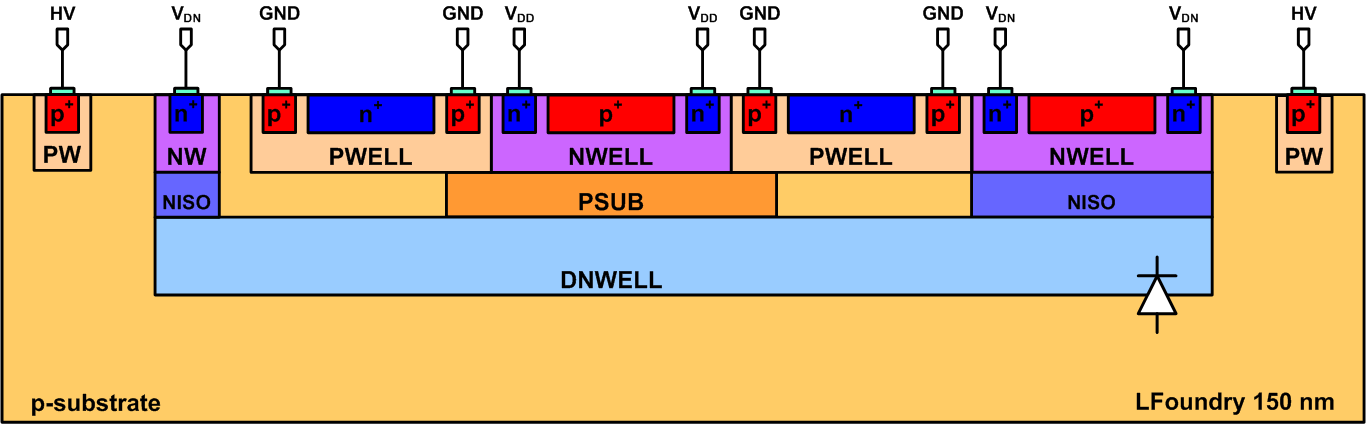}%
\caption{}
\label{fig:pixel}
\end{subfigure}
\caption{Schematics of the RD50-MPW2 sample: (a) chip with marked $8\times8$ active pixel matrix; (b) pixel cross section showing nested p- and n-wells containing readout electronics.}%
\label{fig:pixel_cross_section}%
\end{figure}

\begin{figure}%
\centering
\begin{subfigure}[b]{0.95\textwidth}
\includegraphics[width=\columnwidth]{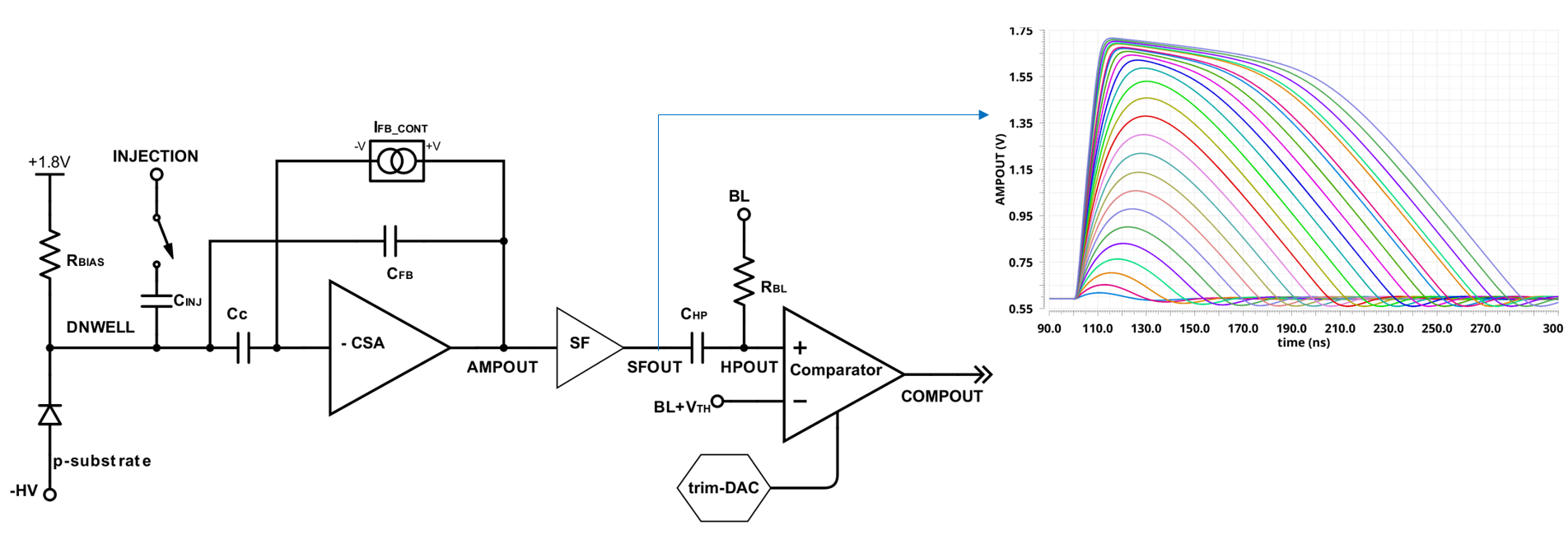}%
\caption{}
\label{fig:pixel_cr}
\end{subfigure}
\\
\begin{subfigure}[b]{0.95\textwidth}
\includegraphics[width=\columnwidth]{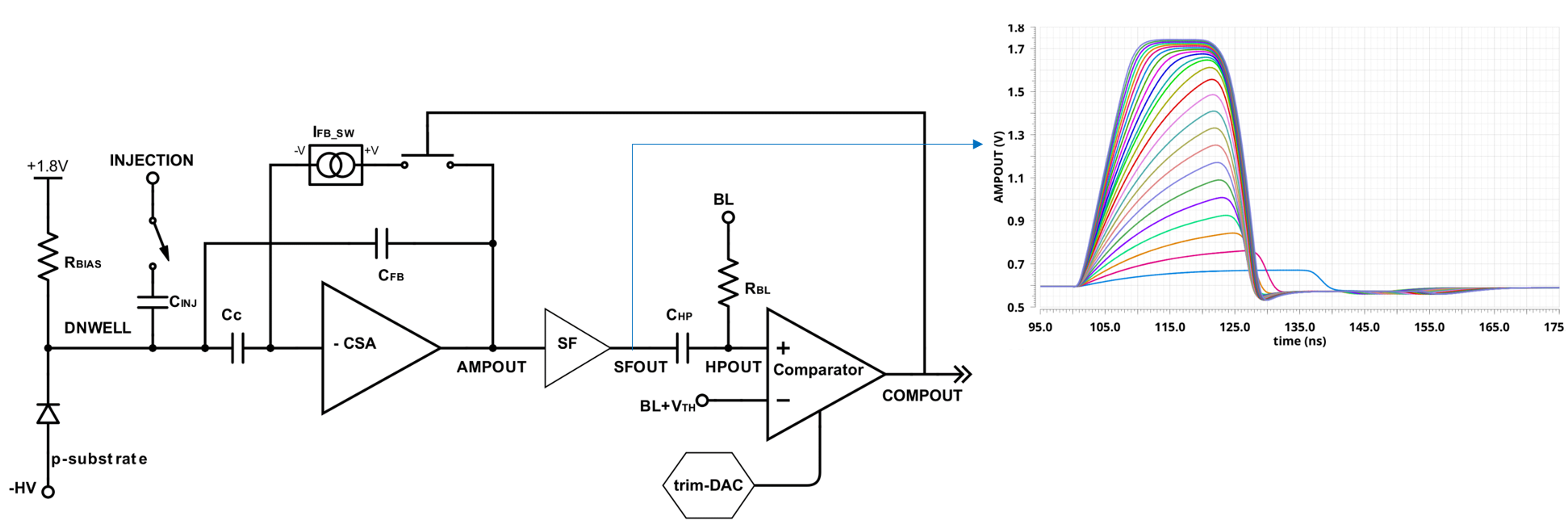}%
\caption{}
\label{fig:pixel_sr}
\end{subfigure}
\caption{Schematics of the in-pixel amplifier and comparator circuit in (a) continuous reset pixel and (b) switched reset pixel. Simulated analogue pulses after the source follower (SF) for input charge between $1\up{ke}^-$ and $25\up{ke}^-$ illustrate the difference between the pixel flavours.}%
\label{fig:pixel_circuit}%
\end{figure}

The chip houses several test structures including an $8\times8$ active pixel matrix with a pixel size of $60\mum\times60\mum$. Each pixel contains a charge sensitive amplifier and a discriminator circuit integrated within several nested p- and n-wells embedded in an n-type collection electrode (Figure \ref{fig:pixel_cross_section}). 
The schematics of the in-pixel circuit is shown in Figure \ref{fig:pixel_circuit}. The pixels come in two variants with different reset mechanisms for discharging the feedback capacitance -- the first four matrix columns, called continuous reset, use a constant current source, while the other four columns use a transistor switch which resets the inputs much faster (called switched reset). In continuous reset pixels the duration of the comparator output signal (\textit{time over threshold} -- ToT) scales with the signal size and can be used to measure the amount of input charge. For this reason the continuous reset type was selected for this study.
Both pixel types also contain a so called calibration circuit with a node for injecting test charge directly into the amplifier input via a MOS capacitor with a capacitance of $C_\mathrm{inj} = 2.8\up{fF}$. The pixel discriminator output is routed out of the chip via an analogue multiplexer allowing reading out one pixel output at a time. 

The chip is configured with several global DACs as well as an independent four bit threshold tuning DAC in each pixel. The configuration and DAQ system for the chip is based on Caribou DAQ platform \cite{caribou} controlled by an FPGA evaluation kit (Xilinx ZC706). 
Discriminator output signals can be recorded either with the Caribou setup or with an external oscilloscope via an intermediate buffer circuit. 

This study was made with two samples: one unirradiated and one irradiated with neutrons at JSI TRIGA reactor in Ljubljana to an equivalent fluence of $5\e{14}\neqcm$ and a reactor background total ionising dose (TID) of $5\up{kGy}$ \cite{snoj, ambrozic}.

\section{Front end threshold, noise and calibration}

\begin{figure}%
\centering 
\begin{subfigure}[b]{0.495\textwidth}
\includegraphics[width=\columnwidth]{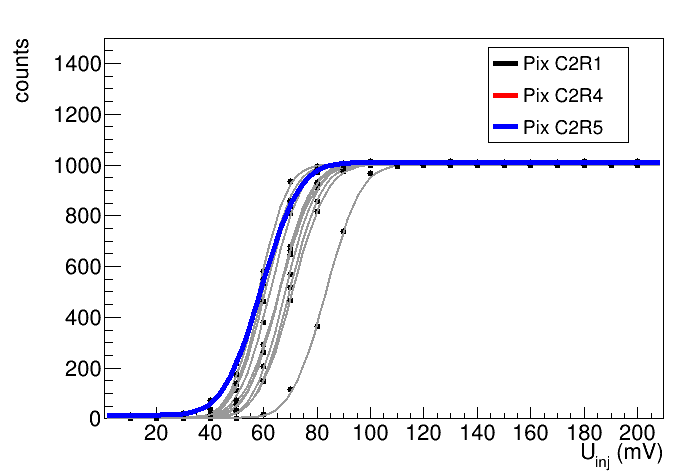}%
\caption{Unirradiated, threshold $950\up{mV}$.}
\label{fig:scurve_0e14_950}
\end{subfigure}
\hfill
\begin{subfigure}[b]{0.495\textwidth}
\includegraphics[width=\columnwidth]{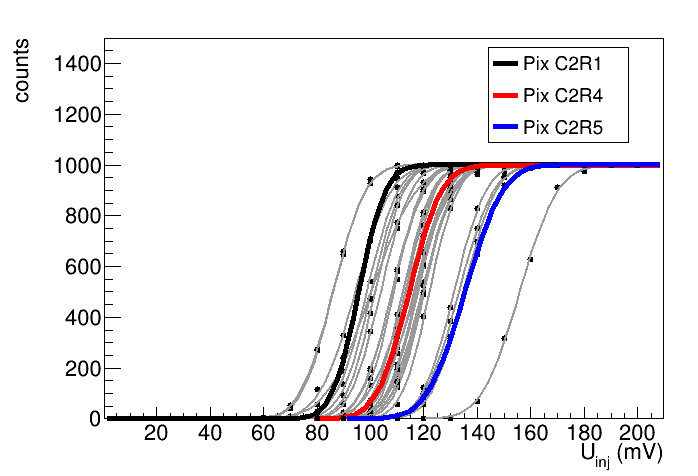}%
\caption{Unirradiated, threshold $1000\up{mV}$.}
\label{fig:scurve_0e14_1000}
\end{subfigure}
\begin{subfigure}[b]{0.495\textwidth}
\includegraphics[width=\columnwidth]{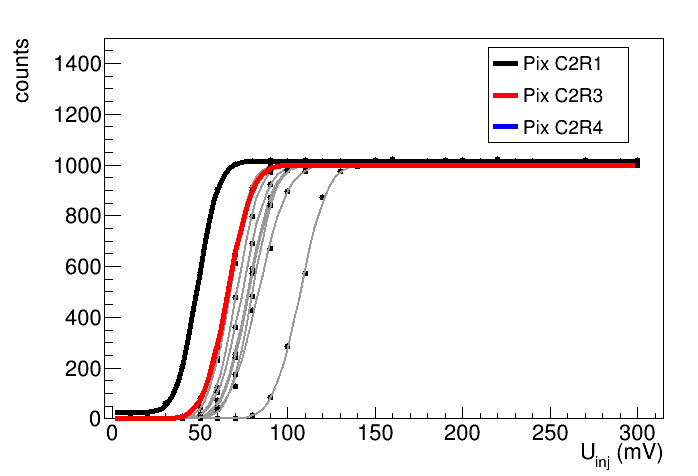}%
\caption{$5\e{14}\neqcm$, threshold $950\up{mV}$.}
\label{fig:scurve_5e14_950}
\end{subfigure}
\hfill
\begin{subfigure}[b]{0.495\textwidth}
\includegraphics[width=\columnwidth]{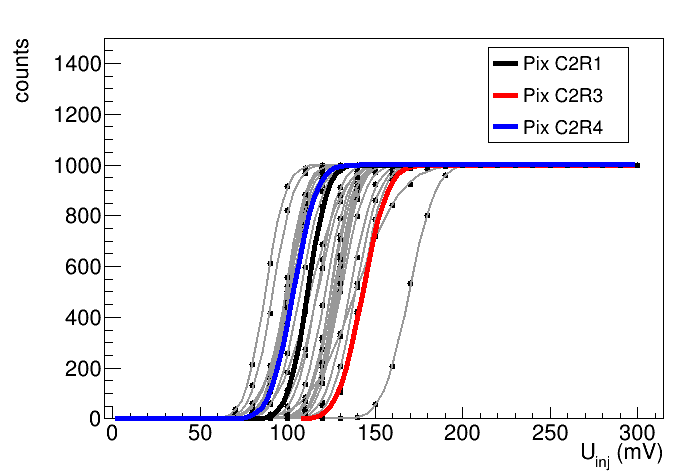}%
\caption{$5\e{14}\neqcm$, threshold $1000\up{mV}$.}%
\label{fig:scurve_5e14_1000}
\end{subfigure}
\caption{S-curve measurements on samples before and after irradiation. Data for noisy pixels with more than 1100 counted pulses is not shown. Threshold is set with respect to comparator baseline of $900\up{mV}$. $U_\mathrm{inj}=100\up{mV}$ corresponds to a charge of $1750\el$. Highlighted S-curves indicate pixels which were used in subsequent measurements.}%
\label{fig:scurve}%
\end{figure}

\begin{figure}%
\centering 
\begin{subfigure}[b]{0.495\textwidth}
\includegraphics[width=\columnwidth]{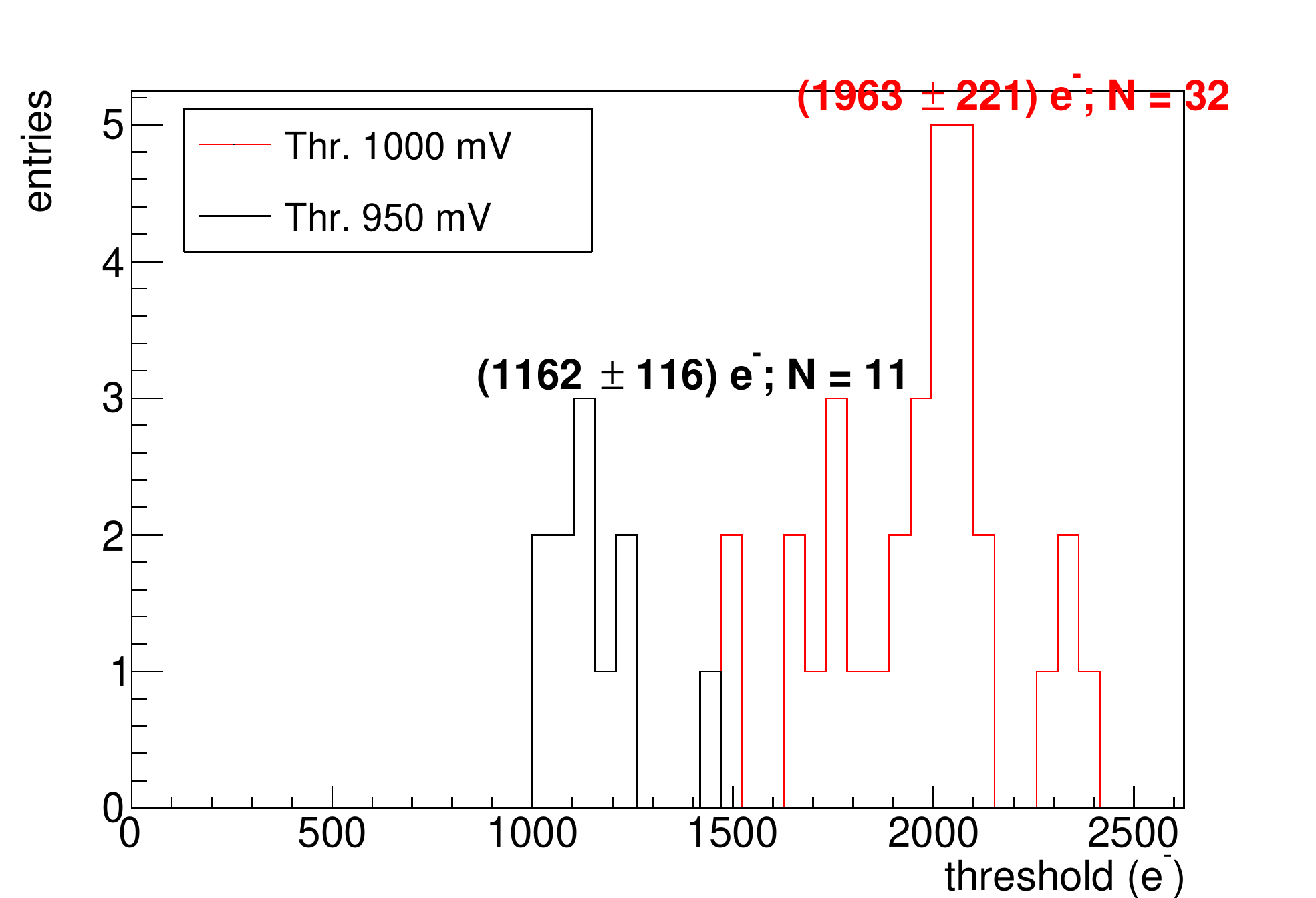}%
\caption{VT50 distribution unirradiated sample.}
\label{fig:scurve_thr_0e14}
\end{subfigure}
\hfill
\begin{subfigure}[b]{0.495\textwidth}
\includegraphics[width=\columnwidth]{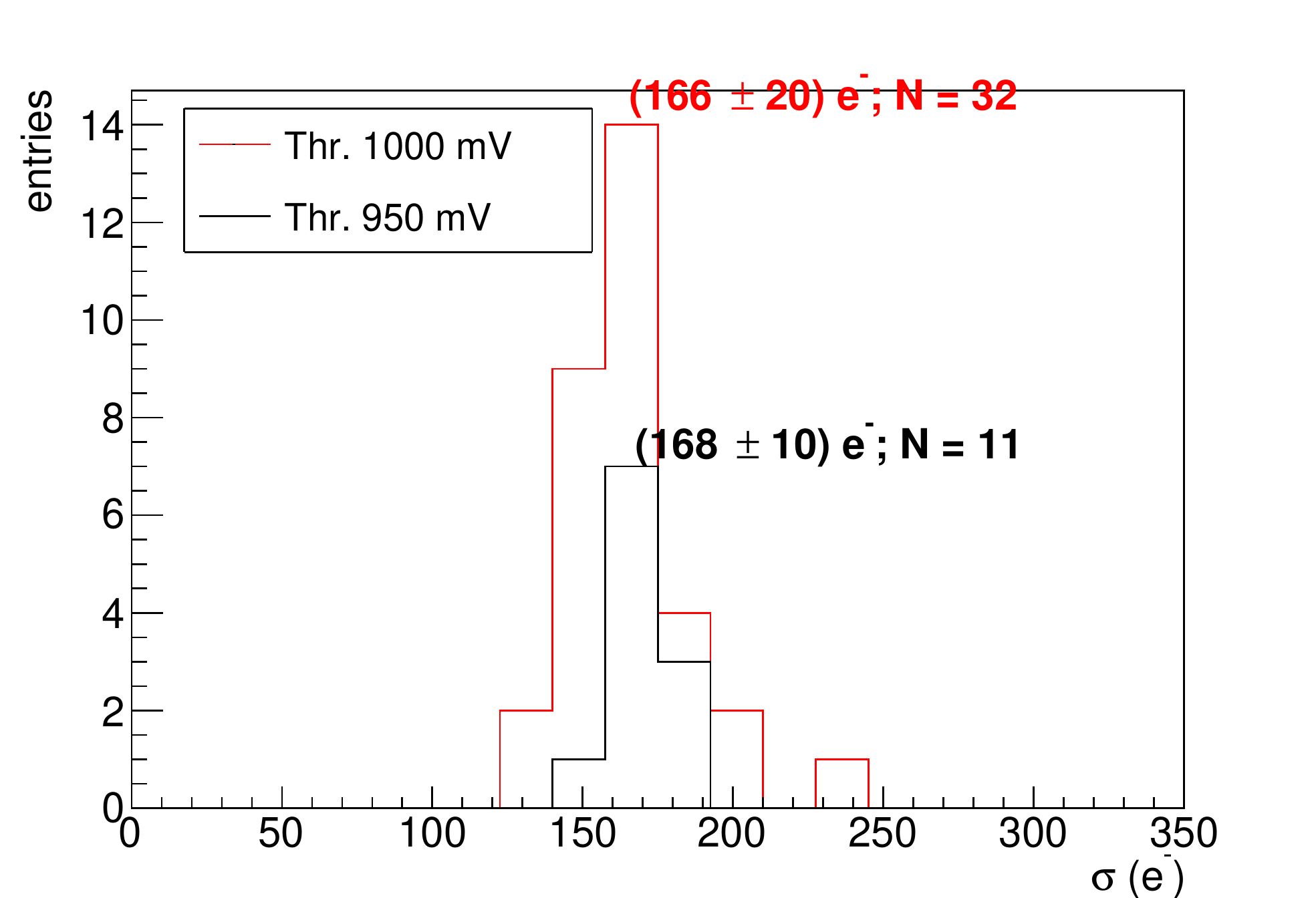}%
\caption{Noise distribution unirradiated sample.}%
\label{fig:scurve_sigma_0e14}
\end{subfigure}
\begin{subfigure}[b]{0.495\textwidth}
\includegraphics[width=\columnwidth]{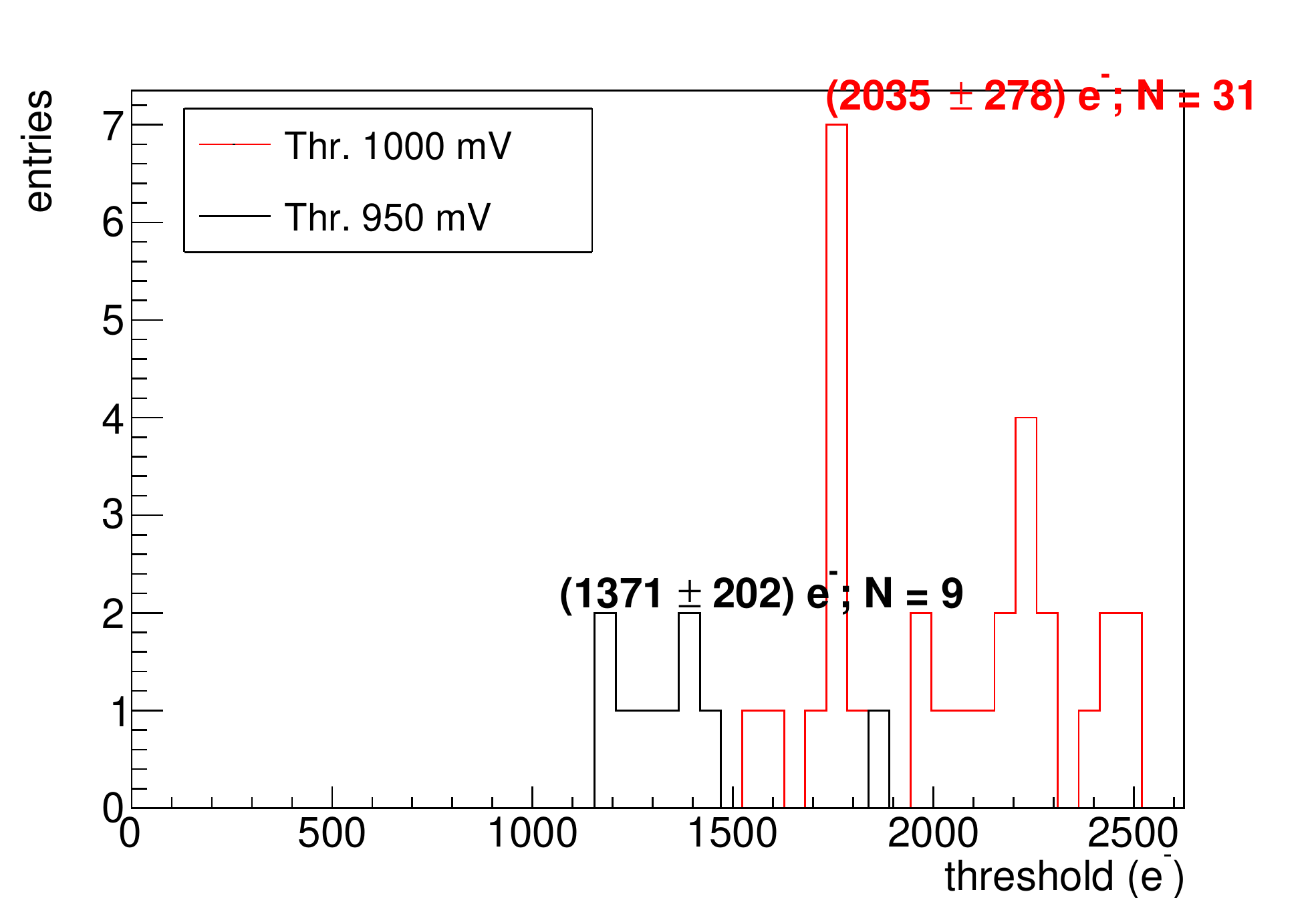}%
\caption{VT50 distribution irradiated sample.}
\label{fig:scurve_thr_5e14}
\end{subfigure}
\hfill
\begin{subfigure}[b]{0.495\textwidth}
\includegraphics[width=\columnwidth]{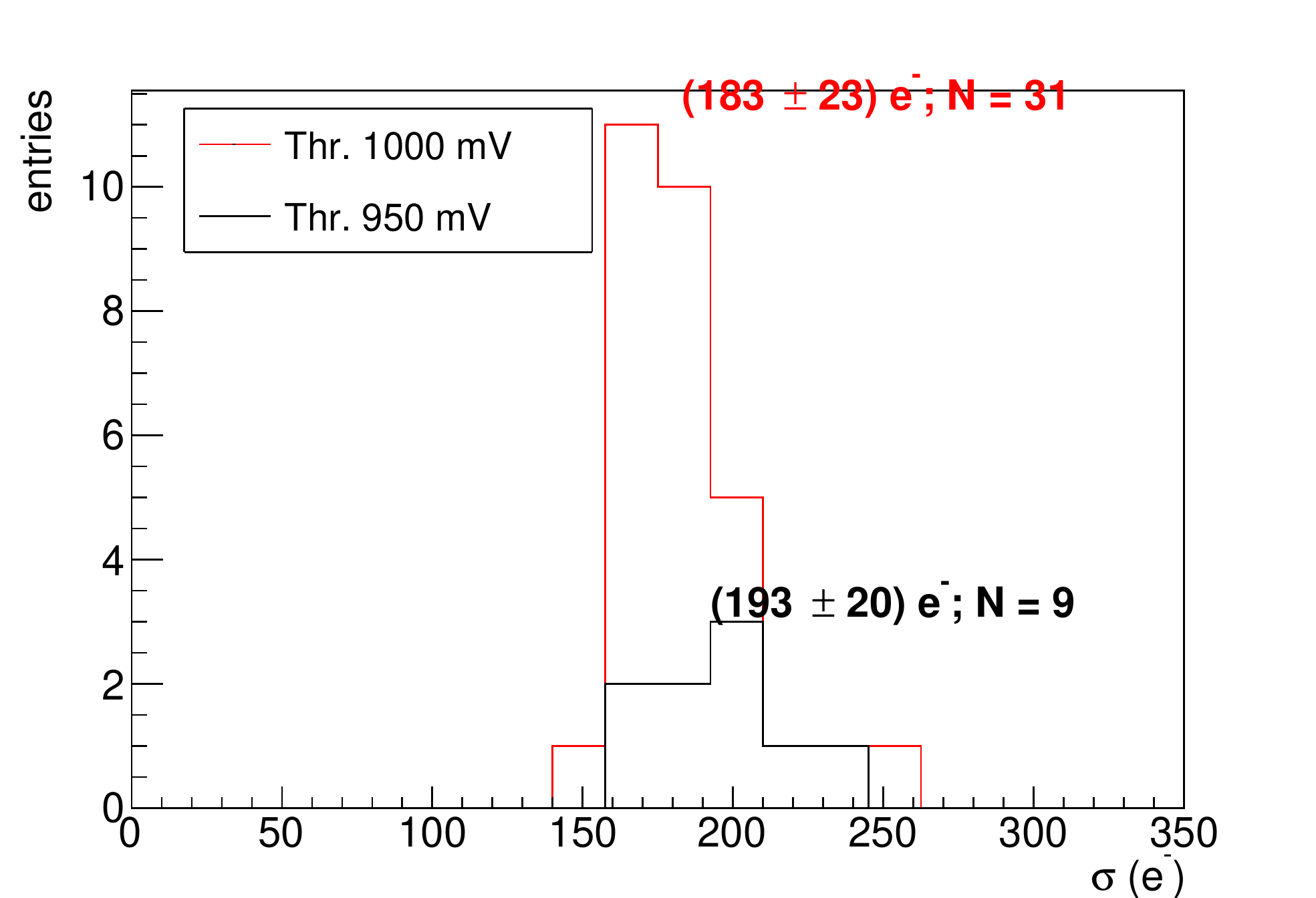}%
\caption{Noise distribution irradiated sample.}%
\label{fig:scurve_sigma_5e14}
\end{subfigure}
\caption{Threshold (VT50) and noise distribution before and after irradiation at $0\up{V}$ bias voltage.}%
\label{fig:distr}%
\end{figure}

Threshold and noise properties of the pixel front end circuit were characterised by analysis of activation curves (S-curves) of the continuous reset pixels using the calibration circuit. The amplifier input was connected via the built in injection capacitance $C_\mathrm{inj}=2.8\up{fF}$ to an external pulse generator providing a voltage step function with variable amplitude $U$. The amount of injected charge $q$ was calculated from the formula $q = C_\mathrm{inj}U$, which evaluates to $17.5\el/\mathrm{mV}$. 
The comparator baseline was set to $900\up{mV}$ and the comparator threshold to two different values of $950\up{mV}$ and $1000\up{mV}$, which are around expected threshold levels from design simulation.
A sweep over a range of injection pulse amplitudes was made with 1000 pulses injected at each amplitude and the number of comparator output pulses was counted with the Caribou and FPGA readout chain. The in-pixel threshold tuning DAC (trim-DAC) was not used in this test and was set to zero. 
All measurements were made at room temperature with the leakage current in the pixel matrix of $20\up{nA}$ and $1\upmu\mathrm{A}$ before and after irradiation respectively at $-100\up{V}$ reverse bias voltage. 
The resulting S-curves for all 32 continuous reset pixels are shown in Figure \ref{fig:scurve}. 
In several pixels the low threshold setting of $950\up{mV}$ was within the noise range, hence a large number of noise hits was detected -- these pixels are excluded from the figure. Measurements were fitted with an error function and the fit parameters sigma (the measure of noise) and the $50\,\%$ point (VT50) were extracted from the fit. The VT50 and noise distributions are shown in Figure \ref{fig:distr}.
The thresholds of $950\up{mV}$ and $1000\up{mV}$ correspond to $1200\el\pm100\el$ and $2000\el\pm200\el$ respectively, with the pixel-to-pixel variance due to manufacturing variations in the front end electronics. These variations have many contributions from different elements of the front end circuit, with the most significant coming from the variance of the feedback capacitance in the charge sensitive amplifier, which has a large influence on the circuit gain. The impact on signal shape is not significant.
The mean noise level is below $200\el$ and does not vary with respect to threshold setting. 
The mean VT50 and noise values increase slightly after irradiation, although statistics is limited. Based on measurements with and without applied biased voltage, around $25\,\%$ of this change is caused by shot noise due to increased leakage current. The remaining contribution is probably due to degradation of front end electronics with irradiation.
For subsequent studies three pixels featuring a low, medium and high VT50 were selected. Their corresponding S-curves are highlighted in Figure \ref{fig:scurve}. The pixels are situated in the third matrix column (column 2) and are surrounded by at least one pixel on each side to avoid any edge effects.

\begin{figure}%
\centering 
\begin{subfigure}[b]{0.495\textwidth}
\includegraphics[width=\columnwidth]{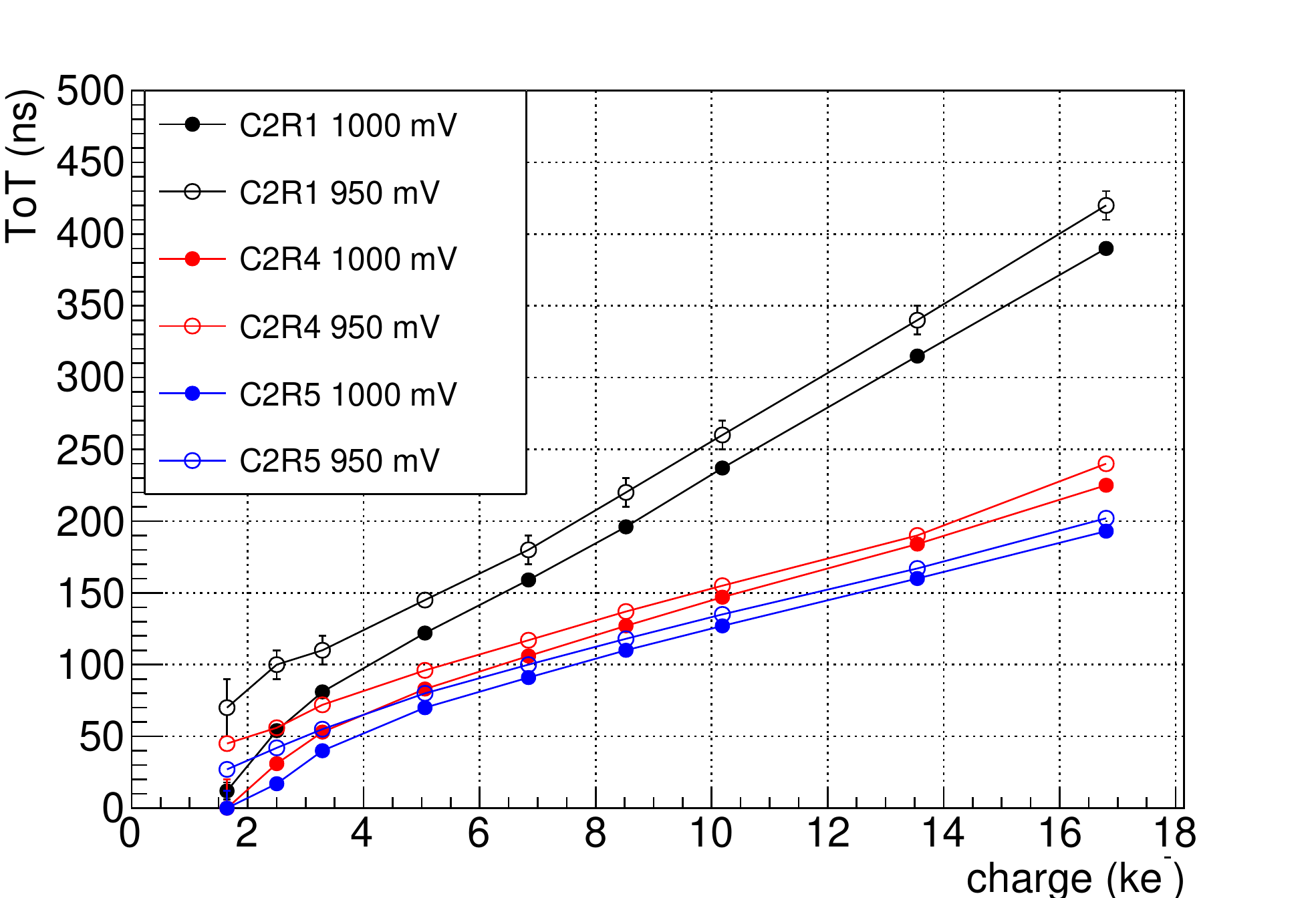}%
\caption{Unirradiated sample.}
\label{fig:calibration_0e14}
\end{subfigure}
\hfill
\begin{subfigure}[b]{0.495\textwidth}
\includegraphics[width=\columnwidth]{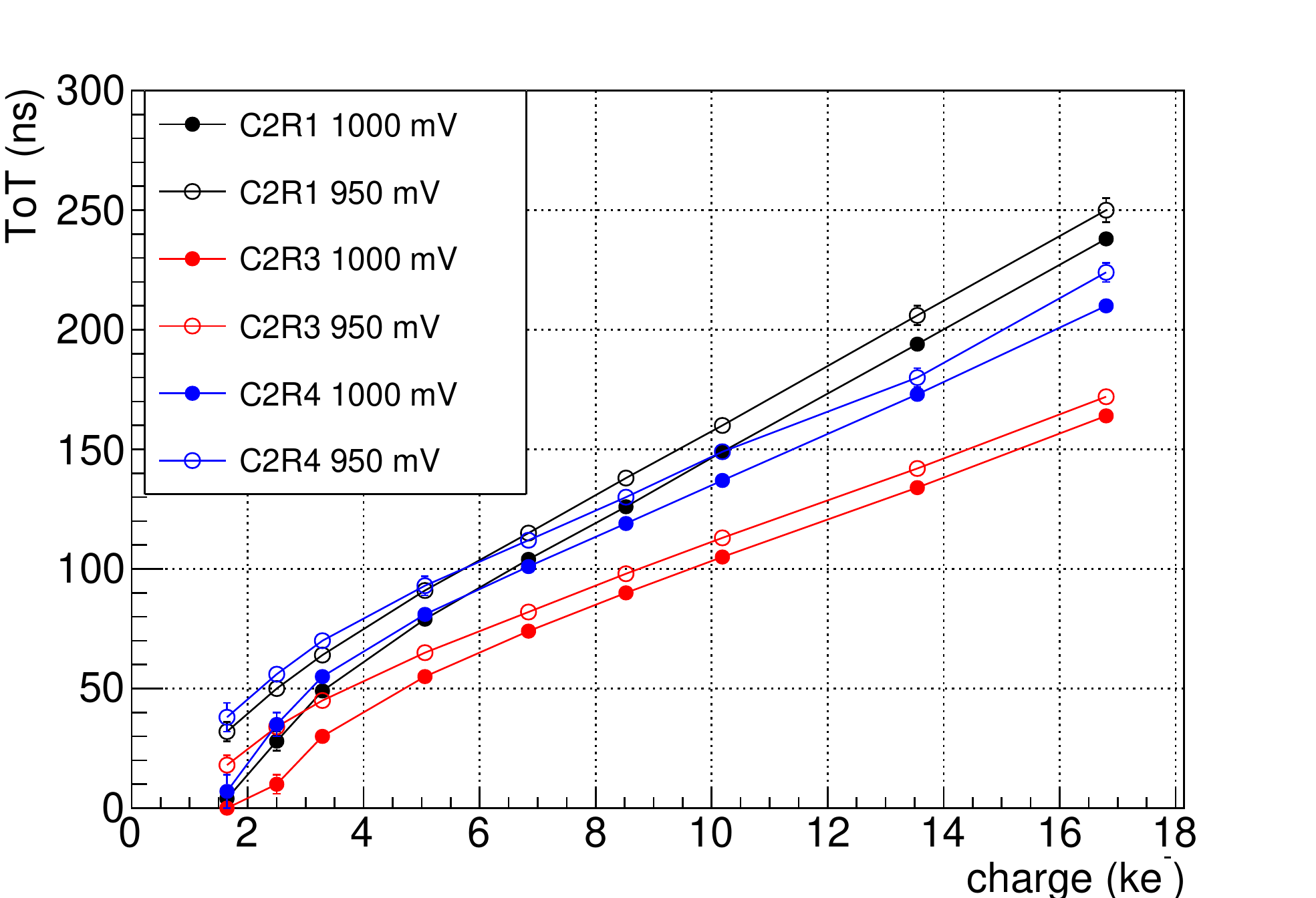}%
\caption{$5\e{14}\neqcm$ irradiated sample.}
\label{fig:calibration_5e14}
\end{subfigure}
\caption{Time over threshold calibration in unirradiated and irradiated sample. 
}%
\label{fig:charge_calibration}%
\end{figure}

The calibration of the comparator pulse duration (ToT) with respect to the amount of injected charge was performed for the selected three pixels. The results are shown in Figure \ref{fig:charge_calibration}. The variation between pixels is significant due to channel to channel variance in gain. 
At lower threshold signals stay above the threshold level longer, resulting in a 5--10$\,\%$ increase in ToT.

\section{Edge-TCT measurement setup}

\begin{figure}%
\centering 
\begin{subfigure}[b]{0.8\textwidth}
\includegraphics[width=\columnwidth]{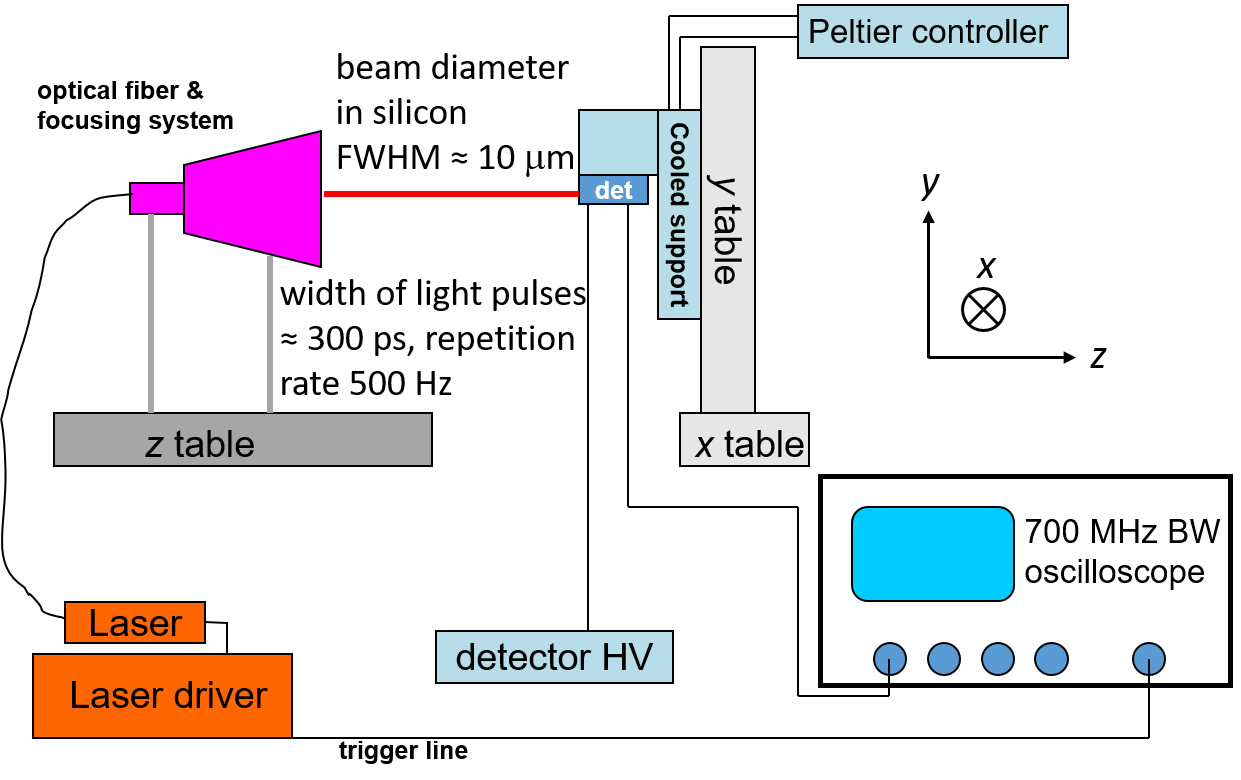}%
\caption{}
\label{fig:tct_setup}
\end{subfigure}
\\
\begin{subfigure}[b]{0.6\textwidth}
\includegraphics[width=\columnwidth]{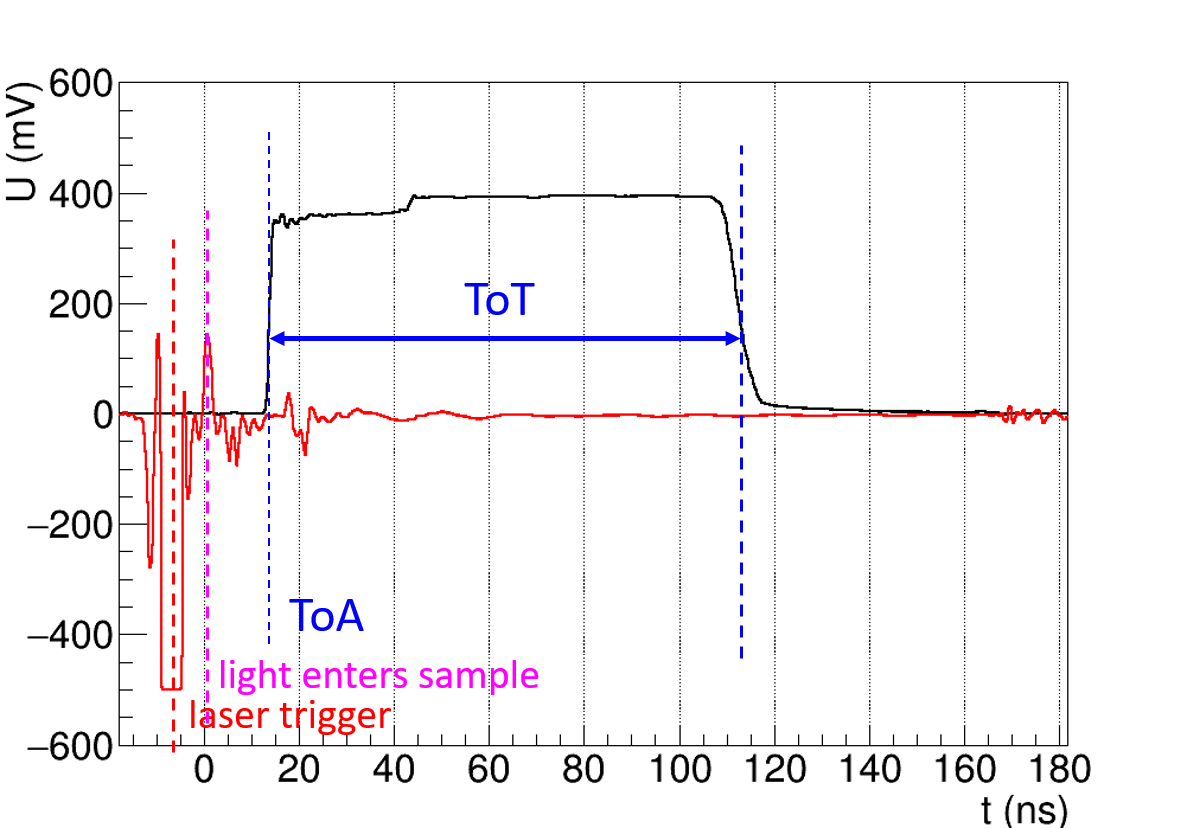}%
\caption{}
\label{fig:tct_pulse}
\end{subfigure}
\caption{Edge-TCT measurement setup: (a) Setup schematics; (b) Typical pulse (black waveform) with indicated time of arrival (ToA) and time over threshold (ToT). Red waveform is the laser trigger pulse. 
}%
\label{fig:tct}%
\end{figure}

Spatially sensitive timing measurements were carried out with the \textit{edge transient current technique} (Edge-TCT) which employs a pulsed laser to generate free charge carriers within the sensor. The schematics of the setup, produced by Particulars\cite{particulars}, is shown in Figure \ref{fig:tct_setup}. Laser beam with a pulse duration of $<1\up{ns}$, a wavelength of $1064\up{nm}$ and corresponding absorption depth of $1\up{mm}$ in silicon is tightly focused to a full width at half maximum of FWHM $<10\mum$ in the beam waist. The sample is mounted on motorized precision placement stages and can be probed with a sub pixel resolution. The sample is oriented in edge configuration with the laser beam entering the silicon bulk from the side, which allows probing the pixel laterally as well as along its depth. 
The edge of the samples was not polished for this measurement.

The signal of free charge carriers drifting within the depletion zone is propagated through the front end and the comparator output is digitised with an oscilloscope. To reduce noise 100 waveforms are averaged at each step and the resulting average waveform is recorded. A typical comparator output pulse from a continuous reset pixel is shown in Figure \ref{fig:tct_pulse}. The nominal output voltage level is in the range of $0\up{V}$ -- $1.8\up{V}$, however when using the $50\,\Omega$ oscilloscope channel termination the pulse is clipped at $400\up{mV}$ due to the limited power of the current drivers in the intermediate buffer. This does not affect the underlying signal shape.
The acquisition is triggered by the trigger output signal from the laser. The relative delay between the trigger signal and the light entering the sample, including delays due to cable length, is $7.5\pm0.5\up{ns}$, which was determined using a passive silicon diode that produces prompt signals. 
The leading edge of the waveform, called \textit{time of arrival} (ToA), is measured at an arbitrarily selected voltage level of $120\up{mV}$ and is used to determine the time walk. The time over threshold (ToT) is used for measuring the signal size. 

\section{Time walk measurements with Edge-TCT}

\begin{figure}%
\centering 
\begin{subfigure}[b]{0.495\textwidth}
\adjincludegraphics[width=\columnwidth, trim={0 0 0 {0.08\height}},clip]{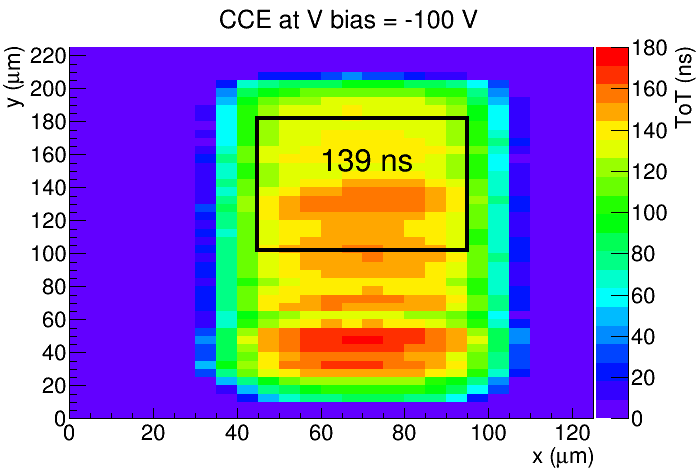}
\caption{Time over threshold unirradiated.}
\label{fig:cce_0e14}
\end{subfigure}
\begin{subfigure}[b]{0.495\textwidth}
\adjincludegraphics[width=\columnwidth, trim={0 0 0 {0.08\height}},clip]{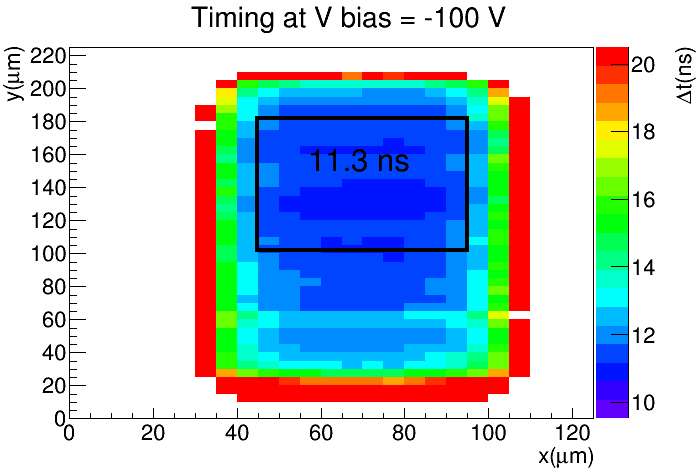}
\caption{Time of arrival unirradiated.}
\label{fig:timing_0e14}
\end{subfigure}
\\
\begin{subfigure}[b]{0.495\textwidth}
\adjincludegraphics[width=\columnwidth, trim={0 0 0 {0.08\height}},clip]{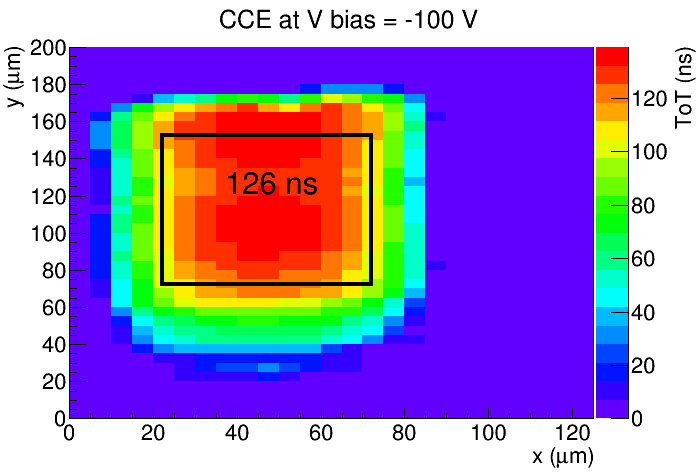}
\caption{Time over threshold irradiated.}
\label{fig:cce}
\end{subfigure}
\begin{subfigure}[b]{0.495\textwidth}
\adjincludegraphics[width=\columnwidth, trim={0 0 0 {0.08\height}},clip]{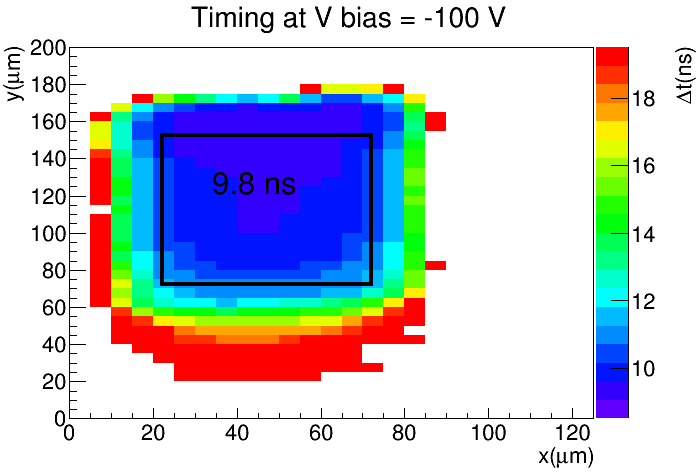}
\caption{Time of arrival irradiated.}
\label{fig:timing}
\end{subfigure}
\caption{Example of Edge-TCT measurements with unirradiated sample (top, pixel C2R1, signal $\approx 6\up{ke}^-$) and irradiated sample (bottom, pixel C2R4, signal $\approx 10\up{ke}^-$): (a, c) Map of collected charge measured in ToT; (b, d) Time of arrival map. Chip surface is at $y=200\mum$ ($170\mum$) and depletion zone grows in negative $y$-direction towards the back side of the sample. The framed area indicates the volume for extracting the average ToT and ToA. }%
\label{fig:etct}%
\end{figure}

Edge-TCT measurements were taken with the sample biased to a voltage of $100\up{V}$ with corresponding depletion depth exceeding $100\mum$. Laser pulse energy was varied in up to five steps to inject different amount of charge, ranging from approximately $2\,\mathrm{ke}^-$ to $10\,\mathrm{ke}^-$ based on ToT calibration from Figure \ref{fig:charge_calibration}.
Space maps of pixel response were obtained by moving the sample in steps of $5\mum$ in $x$- and $y$-directions where the $x$-direction is parallel to the chip surface and $y$-direction is along the sample depth. At each step the collected charge (ToT) and the signal delay (ToA) were measured.
An example of ToT and ToA maps is shown in Figure \ref{fig:etct} which shows a pixel in unirradiated and irradiated sample seen from the side. With unirradiated sample the sensor surface is located at $y=200\mum$ and with irradiated sample at $y=170\mum$. The collected charge distribution in the unirradiated sample shows some modulations along the $y$-direction, which probably come from varying light absorption on chip edge damaged by wafer dicing (the edges were not polished). The time of arrival is uniform on a $1\up{ns}$ level in the entire pixel, except on pixel edges.
The reconstructed width of the pixel is around $60\mum$ which is in agreement with the nominal pixel pitch. The width of the transition region between inefficient and fully efficient volume on pixel edges is dominated by the laser beam spot size of FWHM $\approx10\mum$. This resolution does not allow a precise study of charge sharing on pixel edges. The depth of the depletion zone after irradiation is reduced due to the increased space charge concentration caused by displacement damage.

\begin{figure}%
\centering 
\begin{subfigure}[b]{0.495\textwidth}
\includegraphics[width=\columnwidth]{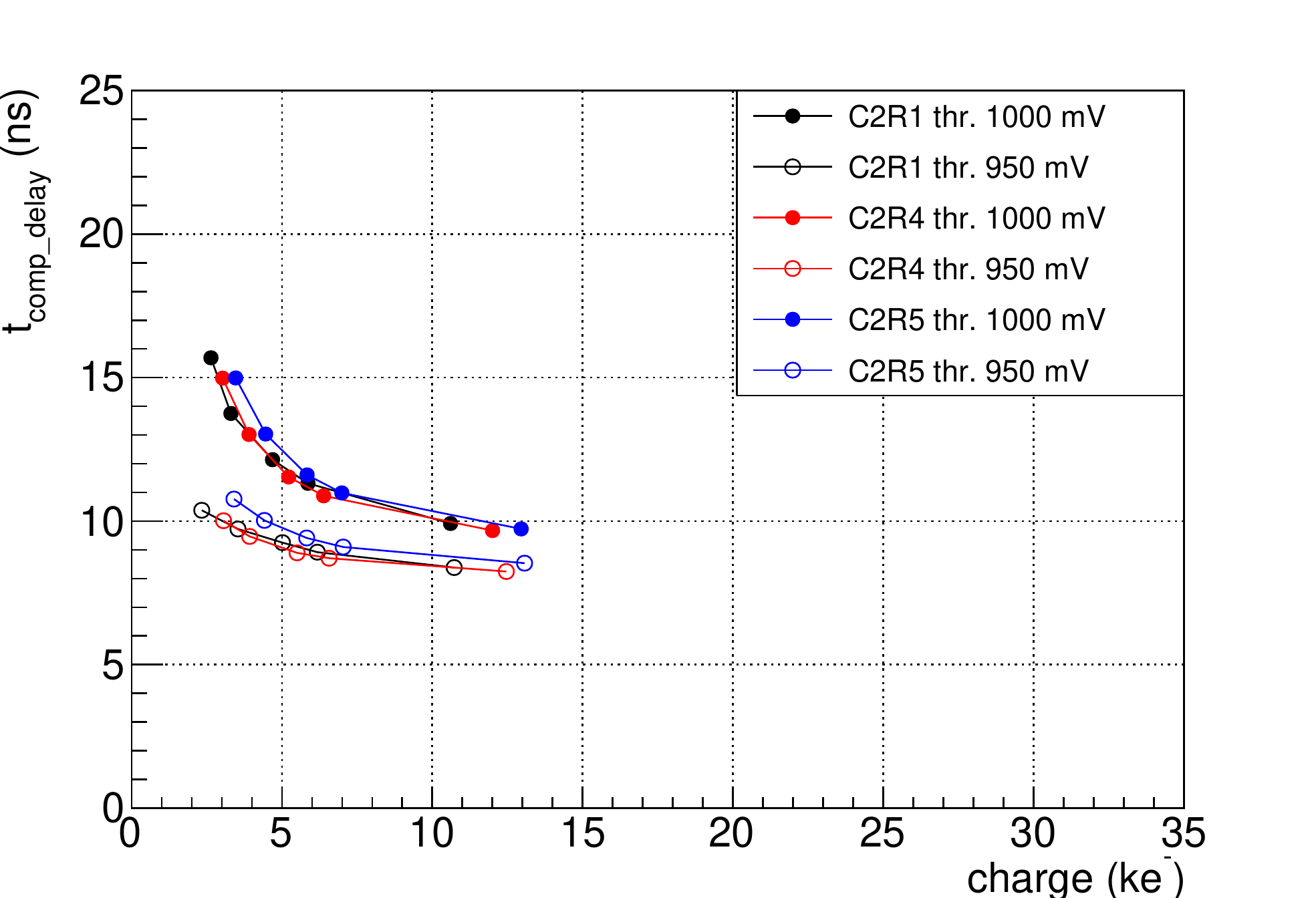}%
\caption{Response delay vs. charge (unirr.).}
\label{fig:timewalk_0e14q}
\end{subfigure}
%
\begin{subfigure}[b]{0.495\textwidth}
\includegraphics[width=\columnwidth]{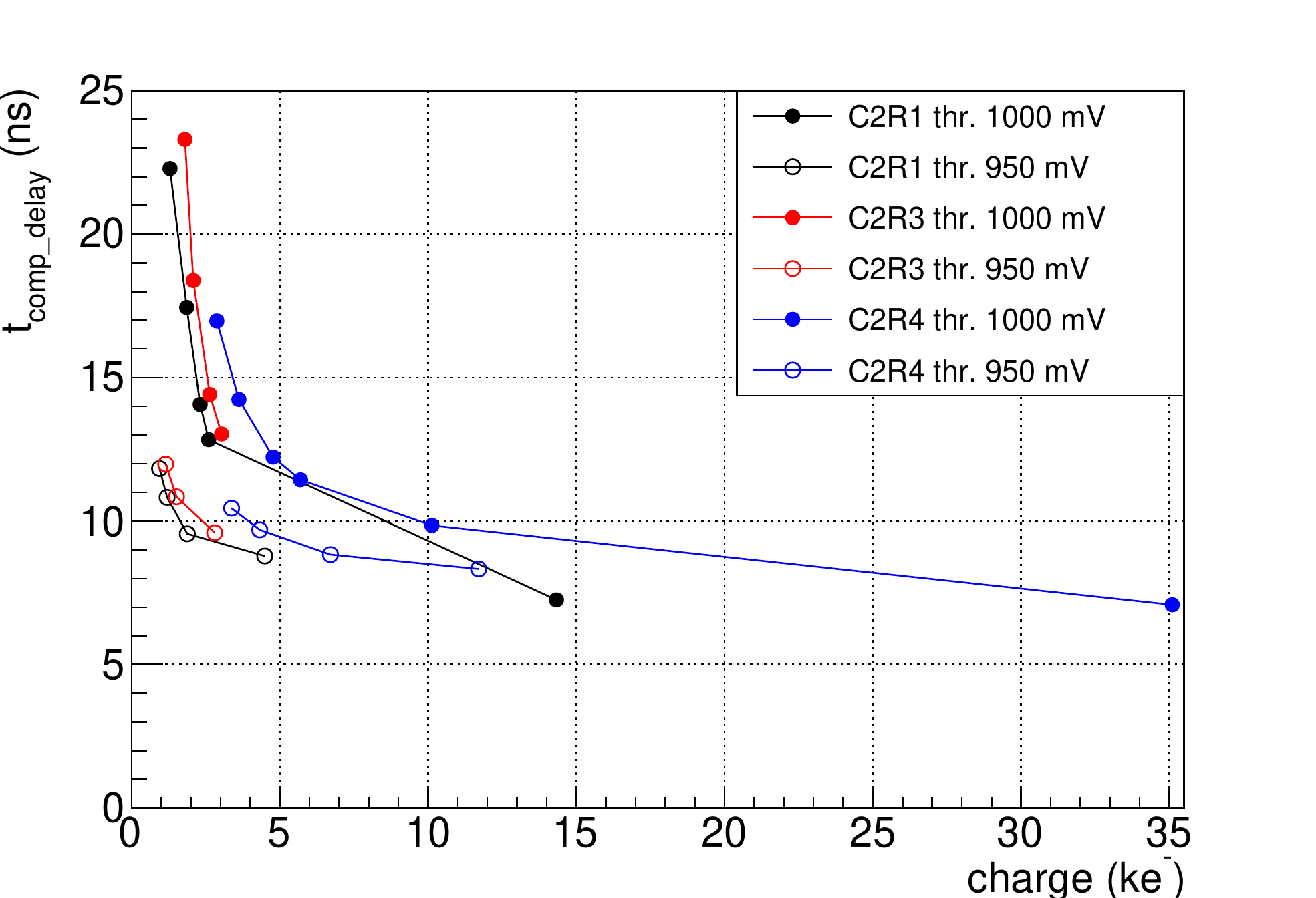}%
\caption{Response delay vs. charge ($5\e{14}\neqcm$).}
\label{fig:timewalk_5e14q}
\end{subfigure}
\caption{Time walk curves measured with Edge-TCT in unirradiated and irradiated sample.}%
\label{fig:timewalk}%
\end{figure}

For evaluation of the time walk ToT and ToA were averaged over a centred volume $50\mum$ wide and $80\mum$ deep starting $20\mum$ below the sensor surface, as indicated in Figure \ref{fig:etct}. These boundaries were selected to contain the most efficient volume of the pixel. This measurement was repeated for each of the three investigated pixels for different laser pulse energies and comparator thresholds. The signal size was converted into electrons using the calibration in Figure \ref{fig:charge_calibration}. The obtained time walk curves are shown in Figure \ref{fig:timewalk}. 

The measured delay ranges from the minimal value of $8\up{ns}$ at a signal size above $\sim 10\,000\el$ (ToT $150\up{ns}$) up to $\approx20\up{ns}$ at signal size of $\sim2\,000\el$ (ToT $20\up{ns}$). 
The time walk curves for different pixels at a fixed threshold coincide relatively well, which indicates that a universal time walk dependence could be applicable for all pixels. 
The delay at a given signal size increases by a few ns with increasing comparator threshold, which is expected, since the level crossing at a higher threshold occurs later.
The agreement between the curves of unirradiated and irradiated sample is also reasonable.

\begin{figure}%
\centering 
\includegraphics[width=0.495\columnwidth]{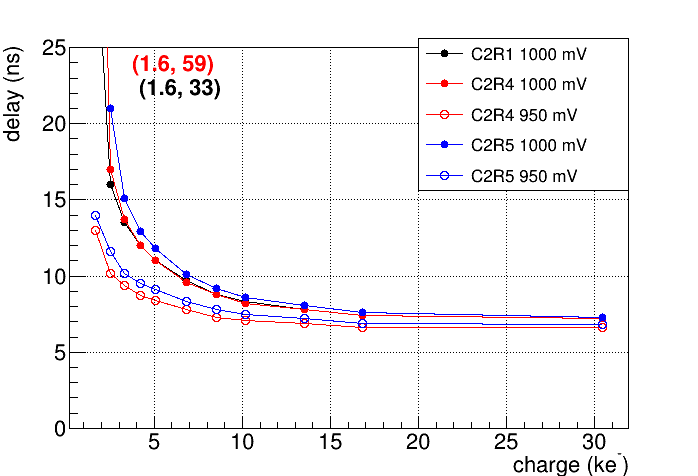}%
\caption{Time walk curve measured with external charge injection (unirradiated sample). Pixel C2R1 could not be operated at low threshold due to noise. }
\label{fig:timewalk_0e14_ext}
\end{figure}

The measurement with the unirradiated sample was additionally validated by measuring the time walk with external charge injection directly into the front end through the injection circuit using a pulse generator. This method bypasses the charge carrier drift in the silicon bulk. The additional latency introduced this way is negligible, since injection circuit is purely passive. The time walk curve shown in Figure \ref{fig:timewalk_0e14_ext} is in a good agreement with Edge-TCT measurements. 
At charge above $10\,000\el$ (ToT $>100\up{ns}$) the delay measured with Edge-TCT is about $1\up{ns}$ longer than with external injection. This difference comes from the charge carrier drift in silicon bulk, which takes place on this time scale.

In practical applications the relevant aspect determining in-time efficiency is the response delay relative to the asymptotic value for the fastest possible response occurring at large collected charge.
Our measurements show that in the irradiated detector the relative output delay is less than $\sim10\up{ns}$ at a deposited charge of $2000\el$, even at a high comparator threshold of $1000\up{mV}$. At the threshold of $950\up{mV}$, which the detector is aiming to operate at, the delay is even smaller. 
With the unirradiated sample the measurements with external charge injection similarly indicate a full in time efficiency for a charge above $2000\el$ at the threshold of $1000\up{mV}$, while at the threshold of $950\up{mV}$ this is true even for a charge of $1600\el$. 
Given that even after irradiation to $5\e{14}\neqcm$ a depletion depth exceeding $100\mum$ can be achieved -- corresponding to a most probable charge deposition of $>8\,000\el$ -- the in-time efficiency in pixel centre is guaranteed. On pixel edges and corners, where charge can be shared by up to four pixels, this is more critical and fine tuning of pixel front end is necessary.

\section{Conclusions}

In the scope of this work tests were carried out with an active CMOS pixel detector RD50-MPW2  before irradiation and after neutron irradiation of $5\e{14}\neqcm$ and TID of $5\up{kGy}$. 
The tests have demonstrated the basic functionality of the active pixel array. External charge injection into the analogue front end has shown all pixels can be operated above noise at a threshold level of $2000\el\pm200\el$, and a noise level below $200\el\pm20\el$.
A new method to evaluate time walk with Edge-TCT was investigated, providing a spatial sensitivity in different parts of the pixel. A uniform output delay was observed in the centre of the pixel. In a 5--10$\mum$ wide band on pixel edges the charge was spread between neighbouring pixels due to a finite beam width of the laser, resulting in reduced signal size and correspondingly longer delay. Delays significantly below $25\up{ns}$ was observed for collected charge above $2000\el$. No significant variations with irradiation were observed. Validation with charge injection through the calibration circuit has generated nearly identical time walk curves, showing that the methods are compatible.

\acknowledgments

This work has been partly performed in the framework of the CERN RD50 collaboration. 
The authors would like to thank the crew at the TRIGA reactor in Ljubljana for help with irradiations of detectors.
The authors acknowledge the financial support from the Slovenian Research Agency (research core funding No. P1-0135 and project ID PR-06802).

\end{document}